\documentclass[oneside,twocolumn,english,prl]{revtex4-1}
 \usepackage{txfonts}
\usepackage[T1]{fontenc}
\usepackage[latin9]{inputenc}
\usepackage{verbatim}
\usepackage{units}
\usepackage{amssymb}
\usepackage{graphicx}
\usepackage{afterpage}

\makeatletter
 
 \@ifundefined{textcolor}{}
 {%
   \definecolor{BLACK}{gray}{0}
   \definecolor{WHITE}{gray}{1}
   \definecolor{RED}{rgb}{1,0,0}
   \definecolor{GREEN}{rgb}{0,1,0}
   \definecolor{BLUE}{rgb}{0,0,1}
   \definecolor{CYAN}{cmyk}{1,0,0,0}
   \definecolor{MAGENTA}{cmyk}{0,1,0,0}
   \definecolor{YELLOW}{cmyk}{0,0,1,0}
 }

\usepackage{graphics}
\usepackage[colorlinks = true,
            linkcolor = blue,
            urlcolor  = blue,
            citecolor = blue,
            anchorcolor = blue]{hyperref}
\usepackage{braket}
\usepackage{epstopdf}

\makeatother

\usepackage{babel}
\begin{document}

\title{Transient coherence of an EIT media under strong phase modulation }

\author{David Shwa}

\author{Nadav Katz}

\affiliation{Racah institute of physics, The Hebrew University, Jerusalem 91904,
Israel}
\begin{abstract}
A strong phase modulated coupling field leads to an amplitude modulation
of a probe field in an electromagnetically induced transparency process.
Time vs. detuning plots for different modulation frequencies reveals
a transition between an adiabatic regime where a series of smooth
pulses are created with a global phase dependent upon the detuning,
and a non-adiabatic regime where a strong transient oscillating response
is added to these pulses. In the extreme non-adiabatic regime, where
the modulation frequency is higher than the transient decay time,
a coherent interference pattern is revealed. Adding a magnetic field
lifts the hyperfine level degeneracy, resulting in a separation of
the original pulse to three pulses. Every pulse has now a different
phase dependent upon the magnetic field causing an interference effect
between the different magnetic level transients. We explore the dynamics
of the magnetic and non magnetic cases and show its resemblance to
the Landau-Zener theory. We also show that combining the global phase
of the pulses with the transient interference allows for a wide magnetic
sensing range without loosing the sensitivity of a single EIT line. 
\end{abstract}
\maketitle
Electromagnetically induced transparency (EIT) is a coherent process,
where a strong coupling field creates a narrow transmission band in
the probe spectrum in an otherwise fully absorptive medium \cite{Fleischhauer2005,Novikova2011}.
The narrow linewidth of EIT makes it suitable for applications in
many fields such as extreme slow light, quantum storage devices, non
linear optics and high sensitivity magnetic sensors. On the other
hand, this narrow linewidth directly limits the bandwidth of data
that can be processed. A signal which has a broader bandwidth than
the EIT linewidth will be absorbed spectrally, or will not be delayed
in the time domain \cite{Tidstrom2007}. In terms of magnetic sensing
this means that although a very high sensitivity is possible using
EIT, it is a problem to probe with this sensitivity a broadband field.
A possible solution to this problem may arise from two directions.
One idea is to use a multimode EIT system where each EIT line has
still a narrow bandwidth but spreading the signal across many systems
allows for a broader signal. Such systems were devised spatially \cite{Dutton2006}
and spectrally \cite{Yavuz2007,Campbell2009} for larger data capacity
as well as for broadband magnetic sensing \cite{Belfi2007,Vladimirova2009}.
A different approach is to use dynamic EIT where the transient response
may have a much broader bandwidth than steady state EIT. The transient
response of an EIT media to a sudden switching \cite{Greentree2002,Meinert2012,Li1995,Park2004}
as well as for ac magnetic field \cite{Margalit2012} was explored
theoretically and experimentally for various regimes. For a constant
detuning the decay of the transients is dictated by the EIT linewidth,
while the frequency of the transient oscillations equals the two photon
detuning \cite{Park2004,Meinert2012}. The detuning can be larger
than the linewidth leading to an under-damped oscillator response.
In the case of a linear sweep through the resonance the the frequency
of the transient is chirped \cite{Park2004} and behaves similarly
to a Landau-Zener transition \cite{Harshawardhan1997,Kenmoe2013}.
Transients were used also as a magnetic sensing technique to measure
the Earth's magnetic field with $1.5\frac{nT}{\sqrt{Hz}}$ sensitivity
\cite{Lenci2012}. 

In this letter we take both concepts of multimode EIT and transient
EIT and combine them together. Using a full mapping of the transient
response as a function of the detuning we are able to show the transition
from an adiabatic regime to a non-adiabatic regime. We also show the
complex interference pattern that arises when a magnetic field is
applied. In this case an interference between transients from three
Zeeman sub-levels is visible. The problem of three level crossing
in the context of Landau-Zener transitions was addressed both theoretically
and experimentally \cite{Carroll1986,Garraway1997,Gaudreau2011,Ivanov2008,Sun2010,Lin2013}.
Experimentally it was shown using 2 two level system (TLS) defects
that have a similar energy gap inside a Josephson junction, that interference
occurs at certain phases when the junction frequency is swept \cite{Sun2010}.
Here, on the contrary, we have the ability to change the levels energy
gap experimentally implementing the three levels Landau-Zener transition
for variable level detuning. Moreover, we show that this interference
can be useful as a high sensitivity magnetometer combined with broadband
phase modulation sweep in order to achieve a wideband high sensitivity
magnetometer. 

We now describe the effect a phase modulated strong coupling field
has upon the temporal shape of a probe pulse going through an EIT
media. The coupling field can be written as follows:

\begin{equation}
\mathbf{E_{c}}(t)=E_{c0}e^{(i\omega_{c}t-i\phi(t))},\label{eq:coupling field}
\end{equation}

where $E_{c0}$ is the amplitude of the coupling field, $\omega_{c}$
is the optical frequency and $\phi(t)$ is the time dependent phase
modulation. In order to describe the change in the probe field due
to this modulated coupling field the full spatio-temporal Maxwell-Bloch
equations for an EIT system needs to be solved \cite{Kiffner2009}.
In the case of long enough interaction media and perturbative probe
intensity the system can reach a steady state solution. The EIT susceptibility
in this case has only temporal and no spatial dependence. For a susceptibility
with no temporal dependence the probe transmission amplitude, $p(t)$
is given by the convolution of the entering signal, $E_{p}(\omega)$,
and the susceptibility, $\chi(\omega)$, hence $p(t)=\mathcal{F}[E_{p}(\omega)\chi(\omega)].$
In the case of a modulated field the susceptibility is time dependent,
and this convolution is not valid. A possible way of solving this
problem is by taking the spectral decomposition of the susceptibility
$\chi(\omega,t)={\displaystyle \sum_{n=-\infty}^{\infty}}e^{in\Omega t}\chi_{n}(\omega)$.
Now the transmission is just $p(t)={\displaystyle \sum_{n=-\infty}^{\infty}}e^{in\Omega t}\mathcal{F}[E_{p}(\omega)\chi_{n}(\omega)]$
\cite{Kiffner2009}. We use a sinusoidal modulation, hence $\phi(t)=Msin(\Omega_{c}t)$
where $M$ is the modulation depth and $\Omega_{c}=2\pi f_{c}$ is
the modulation frequency. The spectrum of such a modulated field can
be described as a sum of Bessel functions 
\begin{equation}
e^{-i\phi}=\sum_{n=-\infty}^{\infty}e^{in\Omega_{c}t}J_{n}(-M).\label{eq:bessel}
\end{equation}

The spectrum of this modulation has narrow peaks separated by a frequency
$\Omega_{c}$ with a full modulation span of $2M\Omega_{c}$. The
transfer function in this case is just an infinite comb of single
EIT lines \cite{Fleischhauer2005} weighted by Bessel functions: 

\begin{equation}
\chi(t)=\sum_{n=-\infty}^{\infty}i\alpha\frac{J_{n}(-M)e^{in\Omega_{c}t}}{\Gamma-i(\Delta-n\Omega_{c})+\frac{R_{c}^{2}}{\gamma_{12}-i(\delta-n\Omega_{c})}}.\label{eq:susc_no_mag}
\end{equation}

Here $\alpha=\frac{N\mu^{2}}{\epsilon_{0}{\hbar}}$ is the two-level
absorption coefficient, with $N$ the density of the atoms and $\mu$
is the transition dipole moment, $\Gamma$ is the homogeneous decay
rate, $\gamma_{12}$ is the decoherence rate of the two ground states,
$R_{c}=\frac{\mathbf{\mathbf{\mathbf{\mu}}}E_{c0}}{\hbar}$ is the
Rabi frequency of the coupling field which is phase modulated, $\Delta$
is the one photon detuning of the probe field and $\delta$ is the
two photon detuning. 

In the case of the D1 line of warm $^{87}Rb$ vapor with buffer gas
the FWHM EIT linewidth is \cite{Figueroa2008} $\gamma_{EIT}=2(\gamma_{12}+\frac{R_{c}^{2}}{\Gamma_{D}+\Gamma})$
where $\Gamma_{D}$ is the Doppler broadening. This linewidth is usually
a few kHz which is much narrower than the pressure broadened homogeneous
linewidth ($\Gamma\sim100\, MHz$) and the Doppler broadening ($\Gamma_{D}\sim500\, MHz$),
thus the probe two level susceptibility is effectively constant for
the full modulation bandwidth as long as $M\Omega_{c}\ll\Gamma$. 

Applying a magnetic field removes the Zeeman degeneracy and the energy
levels of the hyperfine levels will create a ladder according to the
Zeeman splitting of the two lower levels with a Larmor frequency $\mu_{B}B(g_{F}m_{F}-g_{F'}m_{F'})$.
$B$ here is the magnetic field, $\mu_{B}$ is the Bohr magneton and
$g_{F}$ is the Lande coefficient of the hyperfine level. The EIT
susceptibility will be determined by the Zeeman splitting with several
EIT peaks having a certain phase between them. We can write the transfer
function as follows: 
\begin{eqnarray}
\chi(t) & = & \sum_{n=-\infty}^{\infty}\sum_{m_{F}=-F}^{F}\sum_{m_{F'}=-F'}^{F'}i\alpha\label{eq:mag}\\
 &  & \times\frac{J_{n}(-M)e^{in\Omega_{c}t}}{\Gamma-i(\Delta-n\Omega_{c})+\frac{R_{c}^{2}}{\gamma_{12}-i(\delta-n\Omega_{c}-\mu_{B}B(g_{F}m_{F}-g_{F'}m_{F'}))}}.\nonumber 
\end{eqnarray}

\begin{figure}[h]
\includegraphics{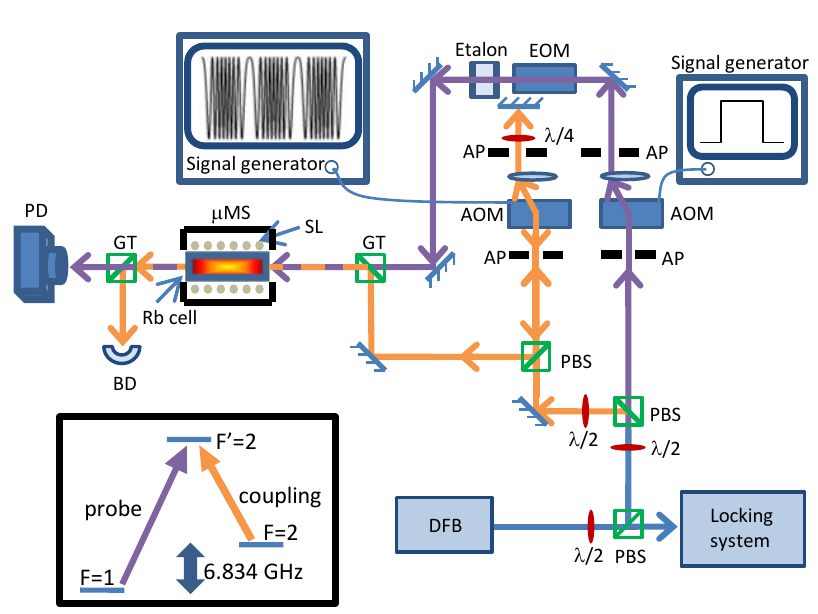}

\caption{\label{fig:setup}The experimental setup. DFB - distributed feedback
laser, PBS - polarizing beam splitter, AP - aperture, AOM - acousto-optic
modulator, EOM - Electro-optic modulator, GT - Glan-Taylor polarizer,
$\mu$MS - $\mu$- metal shield, BD - beam dump, PD - photodiode,
SL - solenoid. }
\end{figure}

The experimental setup is shown in Fig. \ref{fig:setup}. For an EIT
$\Lambda$ scheme we use the hyperfine transitions of The D1 line
of $^{87}Rb$. A DFB laser locked to the $F=2\rightarrow F'=2$ transition
is split into probe and coupling beams using a polarizing beam splitter.
The phase modulation over the coupling field as well as the pulse
creation of the probe is done using acousto-optic modulators. In order
to bring the probe to resonance with the $F=1\rightarrow F'=2$ transition
an Electro-optic modulator is used. The beams (orthogonal polarization)
are combined together using a Glan-Taylor polarizer and pass through
a 7.5 cm cell containing an isotopically pure $^{87}Rb$ with 10 Torr
Ne as buffer gas heated to $\sim40^{\circ}C$. The cell is shielded
from an outside magnetic field using a 3-layers of $\mu$-metal. An
axial magnetic field is created using a uniform solenoid. After the
cell another polarizer is used in order to filter the coupling field
while the probe is detected using an amplified photodiode. 

\begin{figure}[h]
\includegraphics[width=7cm]{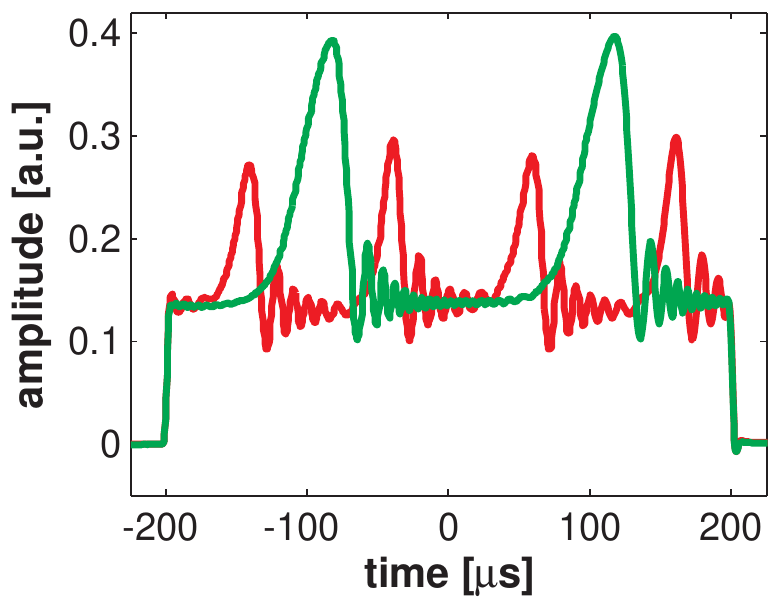}

\caption{\label{fig:ringing}Transient oscillations of the probe amplitude
due to coupling modulation with $f_{c}=5\, kHz$ and $M=20$. Red
- $\delta=0$, green - $\delta=100\, kHz$. }
\end{figure}

Figure \ref{fig:ringing} demonstrates the transmission temporal response
of a square probe pulse with intensity of $0.05\,\frac{mW}{cm^{2}}$
due to a phase modulated coupling field with intensity of $1\,\frac{mW}{cm^{2}}$
in an EIT media. Two major features are observed, one is a train of
pulses that is created with a period and phase that is dependent upon
the coupling field modulation frequency and the detuning \cite{Kiffner2009}.
The second feature is a transient ringing that is associated with
the response of the media to a sudden change in the susceptibility.
This ringing has a chirped frequency as expected \cite{Park2004}.
It decays with a characteristic time that depends upon $\nicefrac{1}{\mbox{\ensuremath{\gamma_{EIT}}}}$
and the chirp rate through the transition \cite{supp1}. The instantaneous
frequency of the coupling field due to the modulation is $\omega(t)=\omega_{c}+\frac{\partial\phi}{\partial t}=\omega_{c}+M\Omega_{c}cos(\Omega_{c}t)$
while the response has the spectral width of the EIT linewidth, thus
the relation between the modulation frequency and the EIT width, sets
the adiabaticity of the response. 

\begin{figure}[h]
\includegraphics[width=8.6cm]{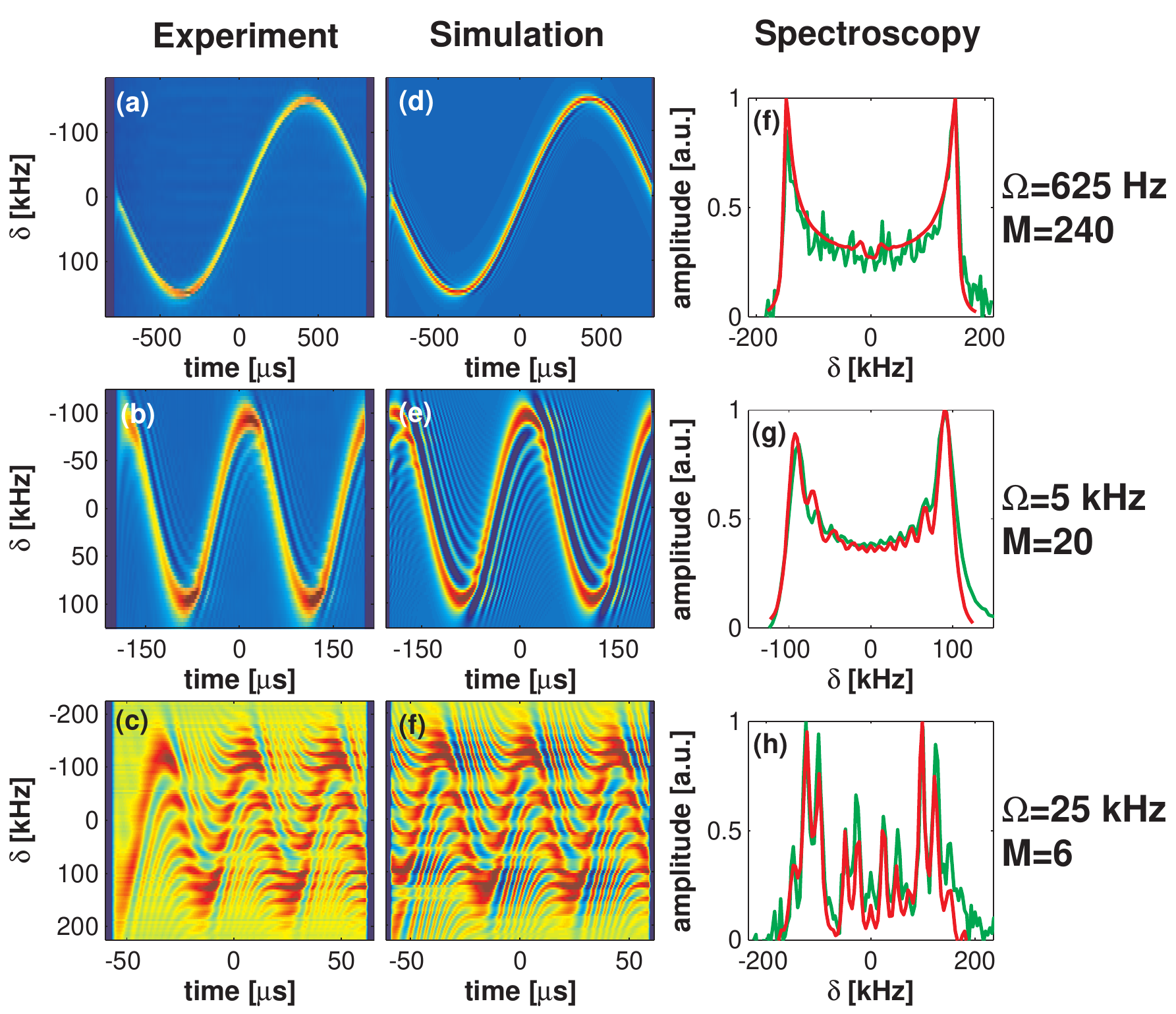}

\caption{\label{fig:spectrum}Adiabatic to non-adiabatic transition. 2D mapping
of the spectro-temporal response of the probe pulse is demonstrated
in the case of (a) adiabatic regime, (b) intermediate regime and (c)
non-adiabatic regime. The parameters of the modulation are (a) $f_{c}=625\, Hz$
and $M=240$, (b) $f_{c}=5\, kHz$ and $M=20$, (c) $f_{c}=25\, kHz$
and $M=6$. All the experiments are done with $\mbox{\ensuremath{\gamma_{EIT}=14\, kHz}}$.
Plots (d-f) show a simulation of the three regimes that takes into
account Eq. \ref{eq:susc_no_mag} with the parameters written above.
Plots (g-i) show the spectrum of the transmission taken as the time
integral for each frequency. Green - experimental spectrum, red -
simulation spectrum. }
\end{figure}

Figure \ref{fig:spectrum} shows experimentally and theoretically
the transition between the adiabatic regime where the modulation frequency
is lower than the EIT linewidth ($\Omega_{c}\ll\gamma_{EIT}$) and
the non-adiabatic regime where $\Omega_{c}\gg\gamma_{EIT}$ . For
both regimes the phase of the pulses is determined by the instantaneous
frequency hence we see a sine like plot as a function of the detuning
with a period $\nicefrac{1}{\Omega_{c}}$ and an amplitude $M\Omega_{c}$.
In the adiabatic regime the transients decay fast enough so they are
hardly noticeable, but as the modulation frequency become comparable
to the EIT linewidth {[}Fig. \ref{fig:spectrum}(b){]} the transient
ringing is clearly observed. In the non-adiabatic regime the modulation
frequency is faster than the decay of the transient ringing creating
an interference between consecutive pulses as can be observed in Fig.
\ref{fig:spectrum}(c). Simulation of these 2D patterns using Eq.
\ref{eq:susc_no_mag} are depicted in Fig. \ref{fig:spectrum}(c-e)
showing a striking similarity to the results. One aspect this linear
response theoretical simulation fails to take into account is the
smearing of the interference pattern when the probe pulse is turned
on as can be visualized particularly in Fig. \ref{fig:spectrum}(c).
The cause of this effect is the gradual build up of the dark state
polariton and consequently the creation of the EIT line that has a
characteristic time of $\nicefrac{1}{2\pi\gamma_{EIT}}$ \cite{Meinert2012}.
Integrating the time domain reveals the steady state spectrum of the
probe light. Figure \ref{fig:spectrum}(f-h) shows the integrated
spectra of the experimental data (green line) as well as the simulation
(red line) in the adiabatic and non-adiabatic regimes. These spectra
fit to the phase modulation spectrum according to the Fourier of Eq.
\ref{eq:bessel}, meaning a delta functions separated by the modulation
frequency, broadended due to the finite EIT linewidth. Another way
of thinking about the interference shown in Fig. \ref{fig:spectrum}(c)
is as a Landau-Zener-Stuckelberg interference pattern where a transition
is crossed repetitively faster than the transient decay time \cite{Shevchenko2010}. 

\begin{figure}[h]
\includegraphics[width=8.6cm]{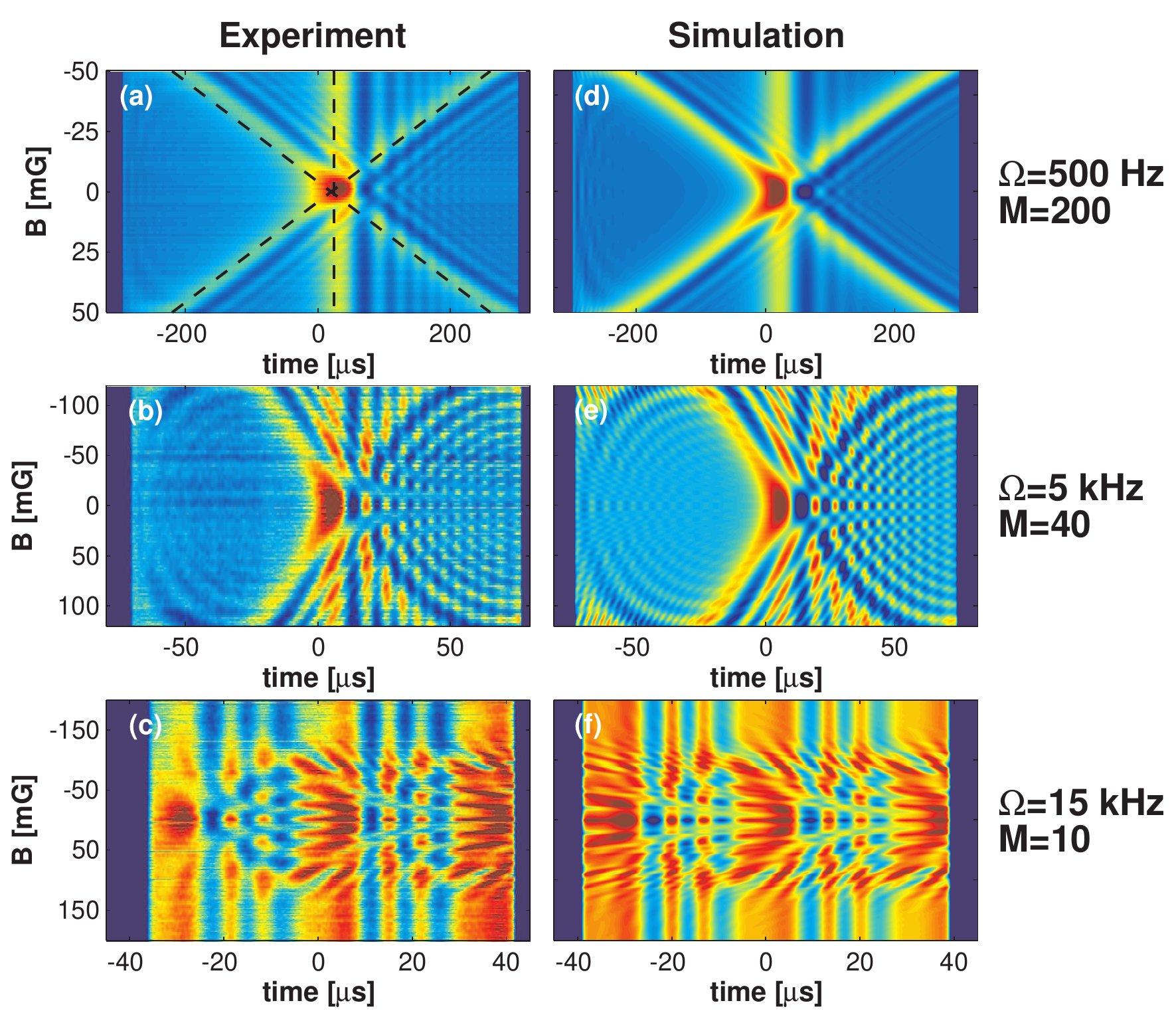}

\caption{\label{fig:magnetic-results}Temporal response of a probe pulse due
to magnetic field for (a) adiabatic regime, (b) intermediate regime
and (c) non-adiabatic regime. The parameters of the modulation are
(a) $f_{c}=500\, Hz$ and $M=200$, (b) $f_{c}=5\, kHz$ and $M=40$,
(c) $f_{c}=15\, kHz$ and $M=10$. Plots (d-f) show a simulation of
the three regimes that takes into account Eq. \ref{eq:mag} with the
parameters written above.}
\end{figure}

Figure \ref{fig:magnetic-results} shows a 2D mapping of the temporal
response of the probe for different magnetic fields (The two photon
detuning is on resonance with the magnetic insensitive transition).
In the adiabatic regime (Fig. \ref{fig:magnetic-results}(a)) it is
possible to see a splitting of the sole pulse in B=0 into three pulses.
These pulses correspond to three EIT lines that are present in the
spectrum. For the D1 line of rubidium, using an arbitrary magnetic
field, up to 7 EIT lines may appear \cite{Cox2011}. Due to the vectorial
nature of the magnetic interaction, the relative strength of these
lines depends on the angle between the beam direction and the magnetic
field as well as the polarization of the pump and probe beams \cite{Yudin2010a}.
The specific configuration we use in our setup is $\mathbf{B\parallel k}$
with linear polarization. In this case only three lines appear in
the spectrum \cite{Yudin2010a,Zibrov2010a} as can be seen experimentally
\cite{Supp2}. The central pulse matches the $\Delta m=0$ and thus
its phase is constant, while the other two pulses correlate with the
$\Delta m=\pm2$, hence having a sinusoidal phase shift. Figures \ref{fig:magnetic-results}(b-c)
show the non-adiabatic regime where every pulse has an oscillating
tail with a certain phase causing an interference pattern. A simulation
based on Eq. \ref{eq:mag} is shown in Fig. \ref{fig:magnetic-results}(d-f)
having the same basic features of the experimental results. 

The observed interference may be understood in the following way.
In the case of hyperfine EIT in a buffer gas and under the condition
of $\gamma_{EIT}\ll\Gamma$ it is possible to treat the two ground
states as a degenerate set of effective two level systems \cite{Vitanov1997,Wu1996}.
Adding a magnetic field removes the degeneracy and splits each two
level system according to the Zeeman frequency \cite{Taichenachev2003}.
In our case due to the selection rules stated above the splitting
is to three groups with $\Delta m=0,\pm2$. As a consequence of this
picture it is possible to think of the magnetic sweep in time as a
chirp of the three TLS's (as depicted by the black dashed lines in
Fig. \ref{fig:magnetic-results}(a)) . Thus, the interference we measure
is a direct consequence of a three Landau-Zener transitions degeneracy
\cite{Carroll1986}. 

The dynamic pattern created by the phase modulation can be used for
broadband magnetic sensing. Each magnetic field has a certain characteristic
pulse timing associated with it. The phase of the first pulse is a
prominent feature for broad magnetic sensing as the total amplitude
of the modulation is $B_{max}=\nicefrac{M\Omega}{\Delta m\mu_{B}g_{F}}$.
This corresponds to $142\, mG$ and $107\, mG$ for \ref{fig:magnetic-results}(b)
and \ref{fig:magnetic-results}(c) respectively. Moreover, the interference
pattern offers a way of measuring accurately the magnetic field in
the area of the interference. The sensitivity to magnetic field is
dependent upon the signal ($S$) to noise ($\widetilde{N}$) ratio
and given by \cite{Belfi2007} 

\begin{equation}
\frac{\Delta B}{\sqrt{\Delta\nu}}=\frac{\sqrt{t}}{S/\widetilde{N}}=\frac{\sqrt{t}\widetilde{N}}{\partial A/\partial B}
\end{equation}

Where $t$ is the measurement time and $\partial A/\partial B$ is
the gradient of the integrated measured amplitude and the magnetic
field. In our case, since the transient ringing is a complex multi
frequency feature the best way to characterize the transients for
different magnetic field is by using the correlation between them.
Using this method we measured the noise and the gradient of the correlation
function and estimate our sensitivity to be $1\frac{nT}{\sqrt{Hz}}$
for the 5 kHz modulation and $0.2\frac{nT}{\sqrt{Hz}}$ in the case
of 15 kHz. The ultimate sensitivity for a given system is \cite{Budker2007}
$\frac{\delta B}{\sqrt{Hz}}=\frac{\hbar}{\mu_{B}g_{F}}\sqrt{\frac{2\pi\gamma_{EIT}}{NV}}$
where $V$ is the volume of the magnetometer. In our case this sensitivity
is $\sim400\frac{fT}{\sqrt{Hz}}$ well below the measured sensitivity.
The main reason for that is the electrical noise in the detector and
amplifier which is used in order to observe the data in the oscilloscope.
Better electronics may allow at least one order of magnitude improvement. 

In conclusion, we show experimentally the transient response of an
EIT media to a phase modulated pump. This response reveals explicitly
the coherent nature of EIT. In the non-adiabatic regime where EIT
peaks are spectrally resolved it is possible to see interference between
the different modes. Albeit the interference between these modes does
not contribute to EIT spectral narrowing and as a consequence to a
slower light propagation \cite{Harris1993,Campbell2009}, it does
create a transient behavior useful for broadband data transfer. Applying
a magnetic field splits the EIT line into three, allowing us to see
an interference pattern caused by the Landau-Zener crossing of the
three EIT lines. Along with the wideband sweep due to the high modulation
index it is shown to be a useful tool for sensitive wideband magnetometry.

\begin{acknowledgments}
We acknowledge helpful discussions with N. Davidson, M. Kiffner and
T. Dey and support of the Bikura (ISF) Grant no. 1567/12.
\end{acknowledgments}
\bibliographystyle{apsrev4-1}

\newpage
\afterpage{\null\newpage}

\section{Supplementary Material}

\renewcommand{\thefigure}{S\arabic{figure}}

\subsection{Landau-Zener representation}

\begin{figure}[h]
\includegraphics[width=8.6cm]{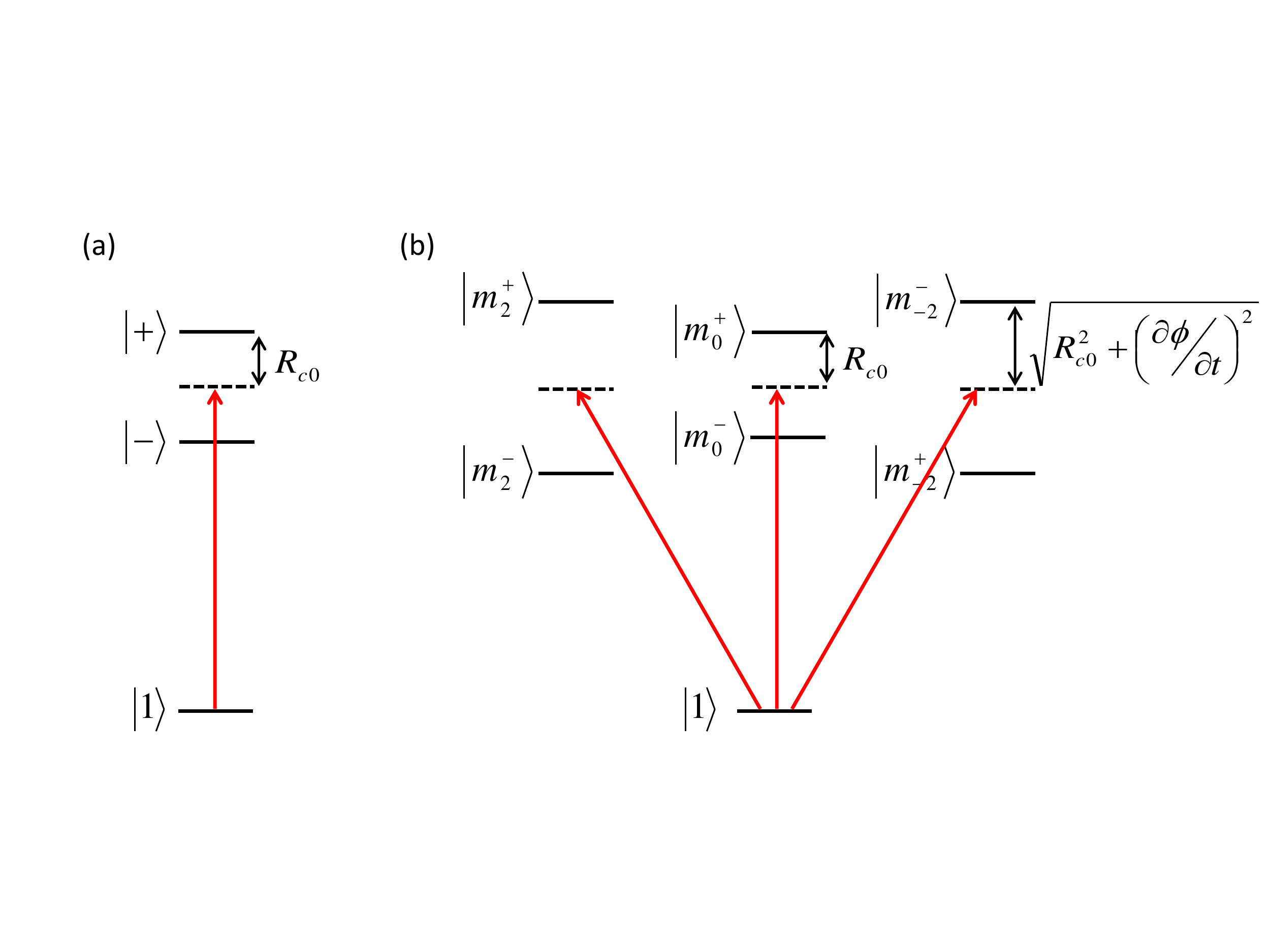}

\caption{\label{fig:dressed}Dressed states for (a) no magnetic field (b) with
magnetic field. Red arrows represent probe transition on two photon
resonance ($\delta=0)$. }
\end{figure}

The ringing observed in Fig. 2 is a manifestation of a non-adiabatic
transition through the EIT resonance. Landau-Zener theory deals with
this kind of transitions and gives analytic prediction to the population
transfer between the levels. In the case of an EIT in buffer gas the
best way to describe the system is using the dressed state picture.
Taking the Hamiltonian of the bare three levels under the rotating
wave approximation 

\begin{equation}
\left(\begin{array}{ccc}
0 & 0 & \frac{1}{2}R_{p}^{*}\\
0 & \delta & \frac{1}{2}R_{c}^{*}(t)\\
\frac{1}{2}R_{p} & \frac{1}{2}R_{c}(t) & \Delta
\end{array}\right)
\end{equation}

where $R_{c}(t)=R_{c0}e^{-i\phi(t)},\, R_{p}$ are the Rabi frequencies
of the coupling and the probe fields respectively, $\delta$ is a
constant two photon detuning and $\Delta$ is one photon detuning
where in the case relevant to us is $0$. In order to see the resemblance
to the Landau-Zener case it is instructive to change to a new basis
where
\begin{equation}
\begin{array}{c}
\Ket{1}'=\Ket{1}\\
\Ket{2}'=\Ket{2}e^{-i\phi(t)}\\
\Ket{3}'=\Ket{3}e^{i\phi(t)}
\end{array}
\end{equation}
 The new Hamiltonian will become 
\begin{equation}
\left(\begin{array}{ccc}
0 & 0 & \frac{1}{2}R_{p}^{*}\\
0 & \delta-\frac{1}{2}\frac{\partial\phi}{\partial t} & \frac{1}{2}R_{c0}^{*}\\
\frac{1}{2}R_{p} & \frac{1}{2}R_{c0} & \frac{1}{2}\frac{\partial\phi}{\partial t}
\end{array}\right)
\end{equation}
 The $2\times2$ matrix of levels $\Ket{2'}$ and $\Ket{3'}$ is a
Landau-Zener Hamiltonian. Under EIT conditions $R_{c}\gg R_{p}$ hence
it is possible to diagonalize this $2\times2$ matrix with two new
dressed levels with eigenvalues $\epsilon_{\pm}=-\frac{1}{2}\delta\pm\frac{1}{2}\sqrt{\delta^{2}+R_{c0}^{2}+(\frac{\partial\phi}{\partial t})^{2}-2\frac{\partial\phi}{\partial t}\delta}$.
In the simple case where $\delta=0$ these states are just $\Ket{+}=\sin\theta\Ket{2'}+\cos\theta\Ket{3'}$
and $\Ket{-}=\cos\theta\Ket{2'}-\sin\theta\Ket{3'}$ with $\tan2\theta=\frac{\frac{\partial\phi}{\partial t}}{R_{c0}}$
\cite{Steck}. This Landau-Zener dynamics is interrogated by the probe
field, meaning that the transition element $\Ket{1}\rightarrow\Ket{3}$
we are measuring in the experiment, carries the dynamics described
above as depicted in Fig. \ref{fig:dressed}(a). In our experiment
a phase modulation sweep in time causes a periodic crossing between
the two dressed levels. 

When a magnetic field is applied the system is split into three sub-systems
with three levels in each on of them as discussed in the main page
and in \cite{Taichenachev2003}. Each one of these sub-systems behaves
exactly as a single EIT system with the exception of a magnetic Zeeman
shift $\mu_{B}B(g_{F}m_{F}-g_{F'}m_{F'})$. As a consequence the energy
levels of the sub-systems $\Delta m=+2$ and $\Delta m=-2$ are reversed
with respect to the magnetic field (with $\delta=0)$ while the energy
levels of the sub-systems $\Delta m=0$ is degenerate up to the interaction
avoided level crossing as depicted in Fig. \ref{fig:dressed}(b). 

\begin{figure}[h]
\includegraphics[width=8.6cm]{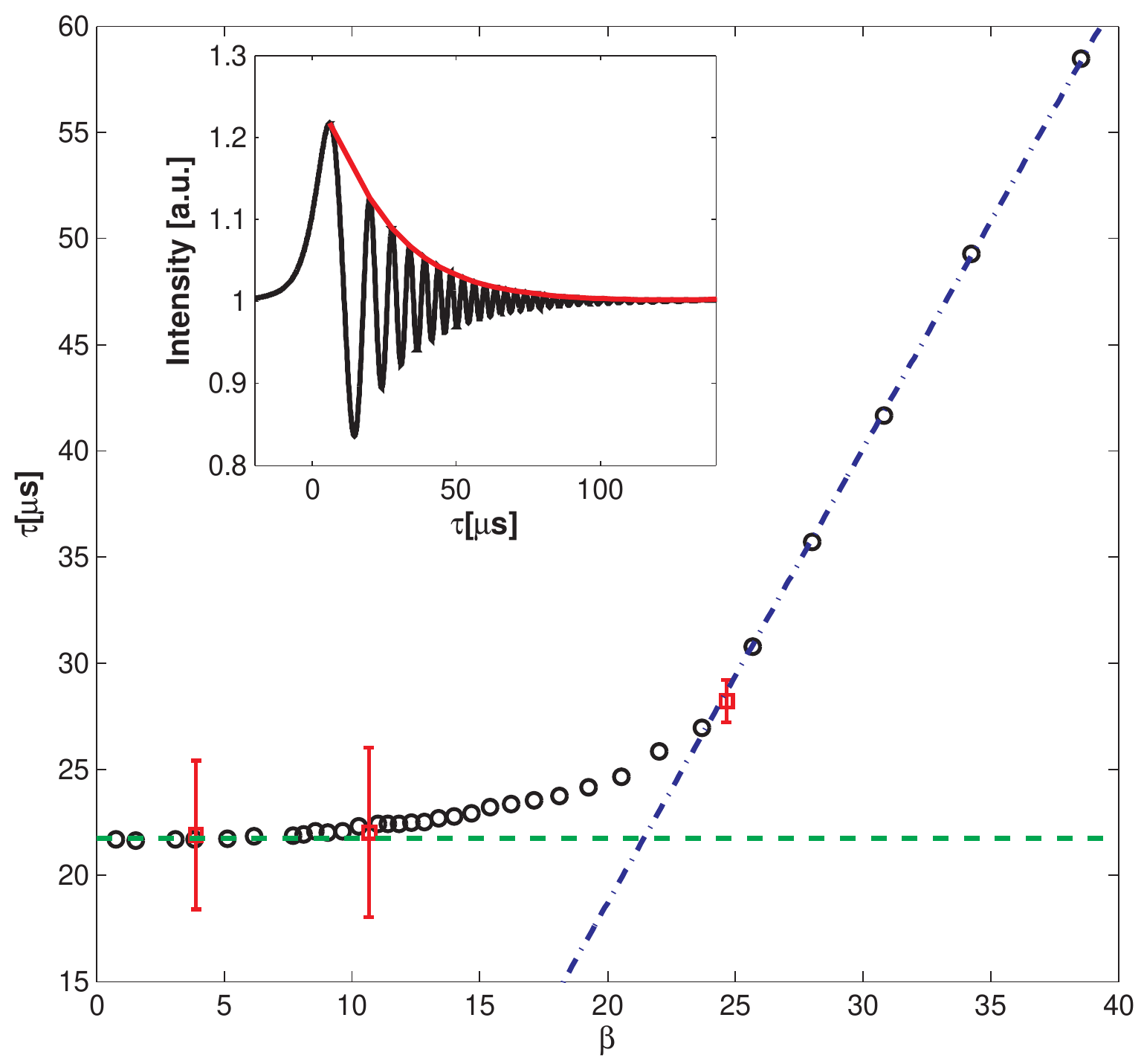}

\caption{\label{fig:decay}Decay time of the ringing at as a function of $\alpha$.
Black circles - simulation, red squares - experiment. The decay time
is calculated using an exponential fit to the peaks of the ringing
as shown in the inset. Green dashed line - the EIT decay according
to $\frac{2}{2\pi\gamma_{EIT}}$. Blue dash dotted line - linear fit
for the adiabatic case. Simulation parameters are similar to the one
in Fig. 3 with variable modulation index and modulation frequency. }
\end{figure}

One interesting characterization of the Landau-Zener transition is
the transition time. This time can be measured by the decay time of
the oscillations after the transition \cite{Vitanov1999}. The two
parameters that determines the transition properties is the coupling
Rabi frequency and the chirp rate defined as $\frac{\partial^{2}\phi}{\partial t^{2}}$.
In the case of sinusoidal phase modulation, where $\phi(t)=Mcos(\Omega t)$,
the chirp rate at $\delta=0$ is $\frac{\partial^{2}\phi}{\partial t^{2}}=M\Omega^{2}$.
It is useful to define the transition using a dimensionless parameter
$\beta=\frac{R_{c}}{\sqrt{\nicefrac{\partial^{2}\phi}{\partial t^{2}}}}=\frac{R_{c}}{\sqrt{M}\Omega}$.
Figure \ref{fig:decay} shows the decay time, $\tau$, as a function
of $\beta$ for our experimental results (red squares) as well as
for our simulation results (black circles). The decay time is found
from an exponential fit to the ringing peaks as depicted in the inset
in Fig. \ref{fig:decay}. It is possible to see that in the diabatic
limit (low $\beta$) the decay time is nearly constant and converging
towards $\frac{2}{2\pi\gamma_{EIT}}$, while at the adiabatic limit
(high $\beta$) the decay is linear with $\beta$. Similar theoretical
results for the Landau-Zener theory have been reported before \cite{Vitanov1999,Mullen1989}. 

\subsection{Temporal dynamics with magnetic field }

\begin{figure}[h]
\includegraphics[width=8.6cm]{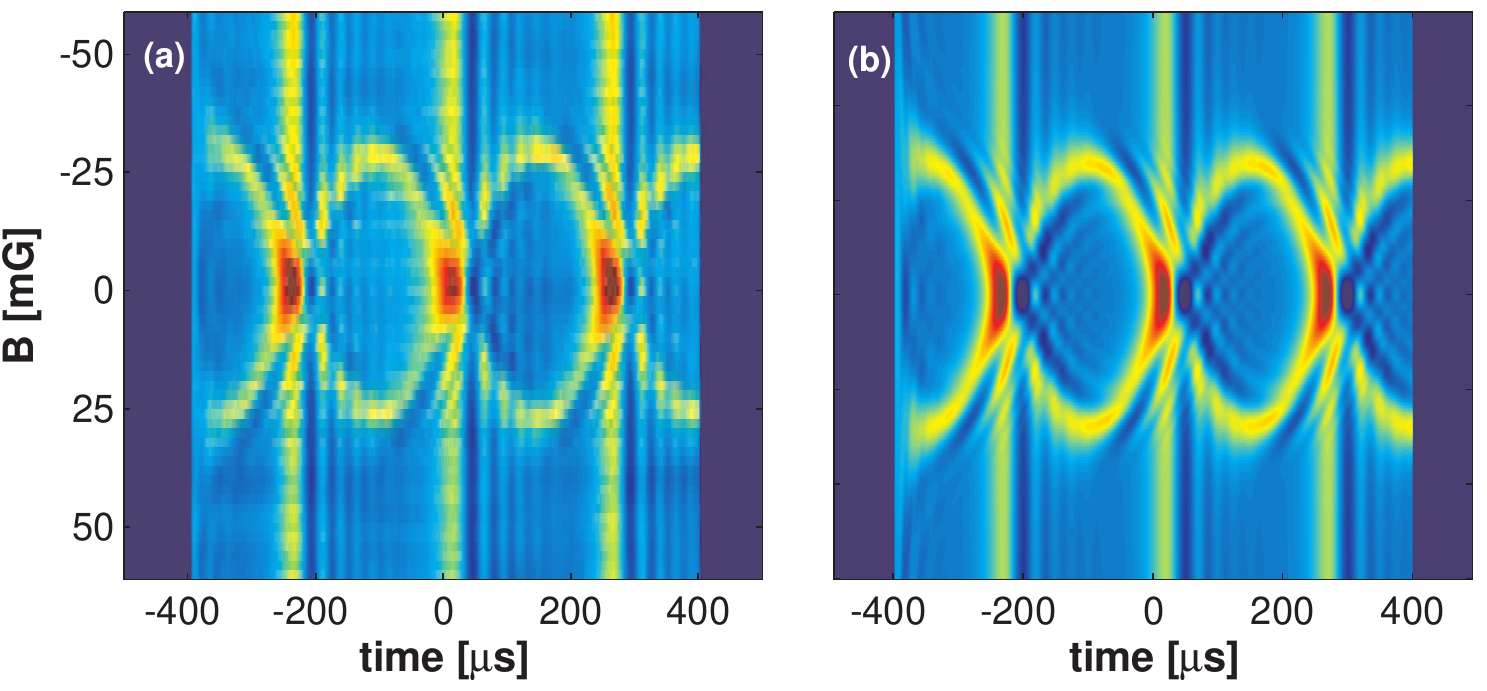}

\caption{\label{fig:broad magnetic}A broad scan both in time and in magnetic
field sweep. Here $f_{c}=2\, kHz$ and $M=10$. (a) Experiment. (b)
simulation. }
\end{figure}

Figure \ref{fig:broad magnetic} shows a broad scan of magnetic field
vs. time. This scan is done in the case of two photon resonance in
the absence of magnetic field. We can distinct clearly the functional
behavior of the three EIT lines for $\Delta m=0,\pm2$. The $\Delta m=0$
line is not dependent upon magnetic field thus its phase is constant
with a pulse every half a cycle. Both $\Delta m=\pm2$ lines are sinusoidally
modulated with a cycle equal to $f_{c}$ and a phase of $\pi$ between
them. Each of these two lines behave exactly like the detuning sweep
of one EIT line (with no magnetic field) under phase modulation (see
for example Fig. 3). This feature is understandable, as applying magnetic
field can be translated to detuning via the Larmor frequency Zeeman
shift. Adding a constant detuning or constant magnetic field creates
a symmetric shift of the two sinusoids until reaching a field larger
than $2M\Omega$. In this case the two sinusoids get separated and
the constant pulse of $\Delta m=0$ disappears. Since the two sinusoids
do not intersect the interference pattern disappears. The major consequence
is that measuring a constant magnetic field accurately using this
method is possible only for magnetic fields with Larmor frequency
smaller than $2M\Omega$.

\subsection{Magnetic field splitting }

\begin{figure}[h]
\includegraphics[width=8.6cm]{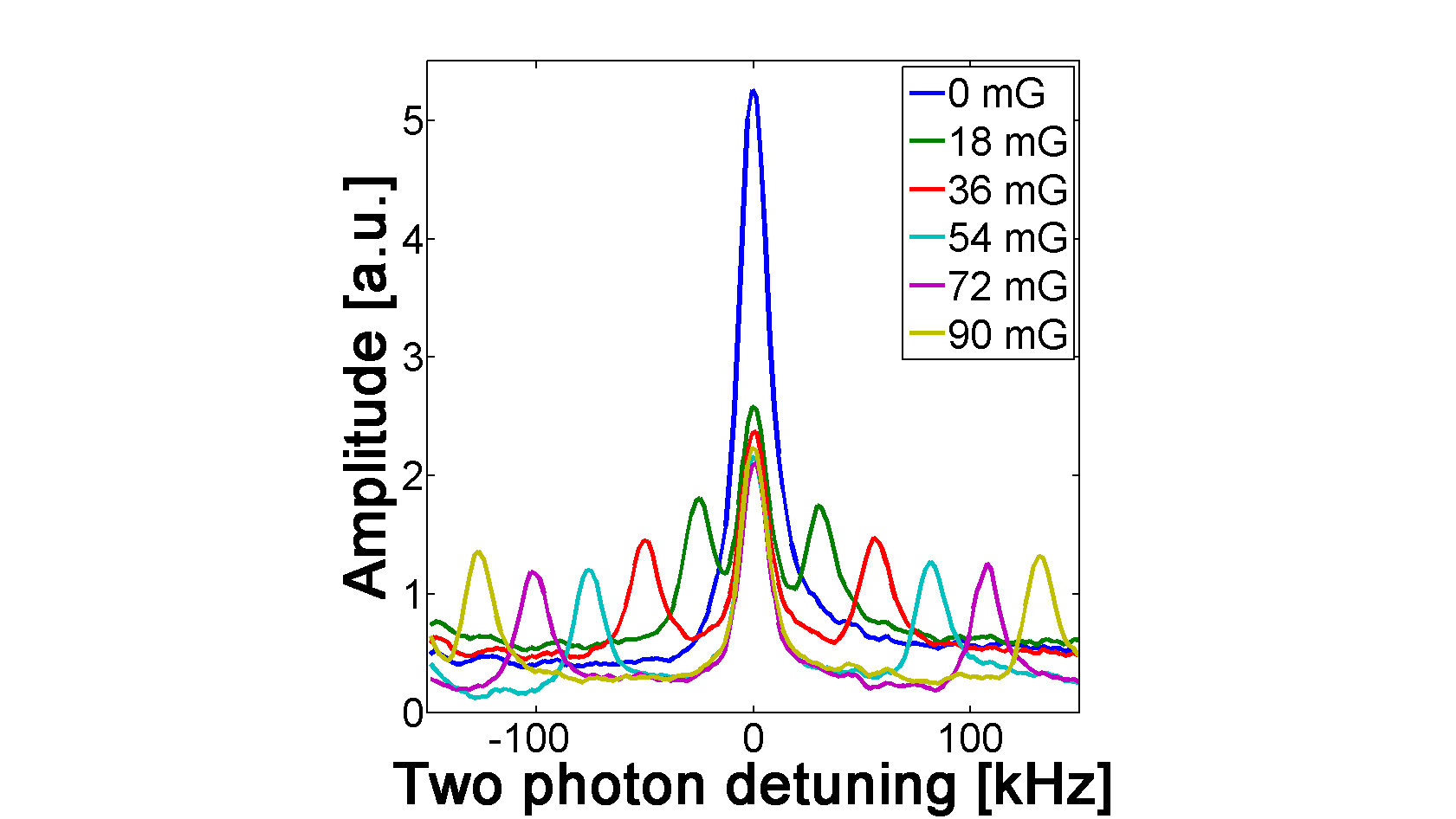}

\caption{\label{fig:Magnetic-splitting}Zeeman splitting of EIT resonance due
to axial magnetic field. The observed spectrum is indeed split to
three levels. }
\end{figure}

As mentioned in the main text, the spectrum of EIT under axial magnetic
field creates a splitting to three sub-levels. This is certainly verified
by the three pulses seen in Fig. 4(a). As a complementary measurement
we also measure the steady state spectrum of the EIT under variable
magnetic field as can be seen in Fig. \ref{fig:Magnetic-splitting}. 

\bibliographystyle{apsrev4-1}

\end{document}